\documentclass[a4paper]{article}
\title{Open string with a background B-field as the first order
mechanics and noncommutativity.}
\author{A.A. Deriglazov\footnote{alexei@fisica.ufjf.br ~ On leave of
absence from Dept. Math. Phys., Tomsk Polytechnical University,
Tomsk, Russia}, ~
C.Neves\footnote{cneves@fisica.ufjf.br} ~
and ~ W. Oliveira\footnote{wilson@fisica.ufjf.br}}
\date{Dept. de F\'\i sica, ICE, Universidade Federal de Juiz de Fora,\\
MG, Brasil.}
\begin{document}
\maketitle
\large
\begin{abstract}
To study noncommutativity properties of the open string with constant
B-field we construct a mechanical action which reproduces classical
dynamics of the string sector under consideration. It allows one to
apply the Dirac quantization procedure for constrained systems in
a direct and unambiguous way. The mechanical action turns out to be the
first order system without taking the strong field limit
$B\longrightarrow\infty$. In particular, it is true for zero mode
of the string
coordinate which means that the noncommutativity is intrinsic property
of this mechanical system. We describe the arbitrariness in the relation existent 
between the mechanical and the string variables and show that 
noncommutativity of the  string variables on the boundary can be removed. 
It is in correspondence with the result
of Seiberg and Witten on relation among noncommutative and ordinary
Yang-Mills theories.
\end{abstract}

{\bf PAC codes:} 0460D, 1130C, 1125 

\newpage

\noindent
{\bf Introduction.} 
Noncommutative Yang-Mills theory arises in a definite limit of string
theory [1]. It has been extracted by Seiberg and Witten  [2] 
starting from the open
string in the presence of a B-field [3-9], with the action for the
corresponding sector being [2]
\begin{eqnarray}\label{1}
S=-\frac{1}{4\pi\alpha^\prime}\int d^2\sigma
\left[\partial_ax^i\partial_ax^i+2\pi\alpha^\prime\epsilon^{ab}
\partial_ax^i\partial_bx^jB_{ij}\right].
\end{eqnarray}
An extremum of the action is supplied by
\begin{eqnarray}\label{2}
(\partial^2_\tau-\partial^2_\sigma)x^i = 0,
\end{eqnarray}
\begin{eqnarray}\label{3}
\lbrace \partial_\sigma x^i+2\pi\alpha^\prime\partial_\tau x^jB_j{}^i\rbrace
\mid_{\sigma=0}^{\sigma=\pi} = 0.
\end{eqnarray}
Open string propagator with the boundary conditions (\ref{3}) contains
an antisymmetric matrix [3, 6], which gives rise to noncommutativity of
the string coordinate on the boundaries [2, 7]. It was suggested [8, 9]
that the noncommutative geometry can be reproduced in the elementary
framework of constrained systems as a result of the Hamiltonian
quantization of the system (\ref{2}), (\ref{3}), by analogy with the
noncommutativity arising for coordinates of a charged particle in
the lowest Landau level [10-12]. To achieve this, there were proposed
rather radical modifications [8, 9] of the Dirac procedure for
constrained systems. There is some discrepancy among the results obtained
in different approaches, which was discussed in [8, 9, 13, 14].

The aim of this work is to quantize the system (\ref{2}), (\ref{3})
following the standard methods [15, 16] without any modifications.
Canonical analysis of this system presents a problem
since the Dirac procedure is initially formulated for the mechanical
system (and then can be generalized for a field with vanishing boundary
conditions). Application of the Dirac formalism to a field with nontrivial
boundary conditions on a compact manifold requires more careful analysis
(see [17] for the open string with the Neumann boundary conditions).
Consistent treatment of such a system implies necessity to represent
the initial dynamics in terms of a mechanical system. Thus we first
rewrite Eqs. (\ref{2}), (\ref{3}) in the form of equations of motion for
mechanical variables $c_n(\tau), ~ n\in Z,$ and then restore the 
mechanical action
which reproduces this dynamics. After that, the Dirac procedure can be
applied to the mechanical system in a direct and unambiguous way. In
particular, $\delta$-function regularization is not necessary in this
case. The last step is to rewrite the results in terms of the string
coordinate $x^i(\tau, \sigma)$, which gives the Hamiltonian formulation 
associated with the theory (\ref{1})-(\ref{3}).

Some of the results thus obtained are as follows.
The corresponding mechanical system 
turns out to be the first order system without taking the
strong field limit $B\longrightarrow\infty$. 
Thus, as a consequence of the mixed boundary conditions, dynamics of 
the mechanical variables is
governed by equations of the first order in time derivative. In
particular,
it is true for the zero modes $c_0^i(\tau)$ of the string coordinate
$x^i(\tau, \sigma)$. As a consequence, $c_0^i$ are canonically conjugated
to each other: $\{c_0^i, c_0^j\}\ne 0$. It means that the noncommutativity
is intrinsic property of this mechanical system. Brackets for the string
coordinates turn out to be noncommutative in the bulk and on the boundary.
We study freedom in relation between the mechanical and the string
variables as well as freedom in choosing brackets which is always
present in the Hamiltonian formalism. This allows one to discuss to what
extent the noncommutativity of the string coordinates can be avoided.
We show that the embedding coordinates of a D-brane can be made
commutative, of course one needs in this case to transform 
simultaneously the Hamiltonian of the system. 
It is in correspondence with the result of Seiberg and Witten on
equivalence of noncommutative and commutative Yang-Mills fields.

\noindent
{\bf Open string with B-field in terms of mechanical variables.} 
To start with, let us continue $x^i(\sigma), ~ \sigma\in [0, \pi ]$ on the interval 
$[0, 2\pi]$ such that $\tilde x^i(\sigma)=x^i(\sigma)$ on $[0, \pi], ~ 
\partial_\sigma\tilde x\mid_\pi=\vec\partial_\sigma x\mid_\pi, \ldots ,$ 
and $\tilde x^i(2\pi)=\tilde x^i(0), ~ \vec\partial_\sigma\tilde x^i
\mid_{2\pi}=\stackrel{\leftarrow}{\partial}_\sigma
\tilde x^i\mid_0, \ldots ~.$ It can be 
further continued on $\sigma\in(-\infty, \infty)$ as a periodic function 
$\tilde x^i(\sigma +2\pi n)=\tilde x^i(\sigma)$. As a result, any 
solution of the problem (\ref{2}), (\ref{3}) can be presented in the form 
\begin{eqnarray}\label{4}
x^i(\tau, \sigma)=c_0^i(\tau)+\sum_{n=1}^\infty\frac 1n
\left[c_n^i(\tau) \cos n\sigma+c_{-n}^i(\tau)\sin n\sigma\right],
\end{eqnarray}
where $c_n^i(\tau), ~ n\in Z$, are our mechanical variables. Substitution 
in Eqs. (\ref{2}), (\ref{3}) gives the equations of motion 
\begin{eqnarray}\label{5}
\ddot c_0^i=0, \qquad \ddot c_n^i+n^2c_n^i=0, \qquad n\ne 0; 
\end{eqnarray}
\begin{eqnarray}\label{6}
\dot c_0^i=0, \qquad \dot c_n^jB_j{}^i+
\frac{n}{2\pi\alpha^\prime}c_{-n}^i=0, 
\qquad n>0,
\end{eqnarray}
from which one finds further consequence
 
\begin{eqnarray}
\label{6.1}
\dot c^i_{-n}- 2\pi\alpha^\prime nc^j_nB_j{}^i=0.
\end{eqnarray} 
\newpage
\par
\noindent Consequently, the complete dynamics can be rewritten in the 
equivalent form as 
\begin{eqnarray}\label{7}
\dot c_0^i&=&0, \cr
\dot c_n^jB_j{}^i+\frac{n}{2\pi\alpha^\prime}c_{-n}^i&=&0, \cr
\dot c_{-n}^i-2\pi\alpha^\prime nc_n^jB_j{}^i&=&0, \qquad n>0,
\end{eqnarray}
which consist of equations of the first order only.
Eq. (\ref{7}) follows from the first order action \footnote{Other 
possibility is to take as independent the equations (\ref{5}),(\ref{6}) 
with $n>0$. Then the action can be chosen in the second order form 
$S=\int d\tau
\left[\frac 12\dot c_0Bc_0+\sum_{n=1}^\infty\frac 12
\dot c_n\dot c_n-\frac{n^2}{2}c_nc_n+(\dot c_nB+
\frac{n}{2\pi\alpha\prime}c_{-n})\lambda_n\right]$.
It looks less natural since it involves the Lagrangian multipliers 
$\lambda_n$.} 
\begin{eqnarray}\label{8}
S_f=&&\int d\tau
\left[\frac 12\dot c_0^iB_{ij}c_0^j+\sum_{n=1}^\infty f_n\left(
\pi\alpha^\prime
\dot c_n^iB_{ij}c_n^j-\right.\right. \cr
&&\left.\left.\frac{1}{4\pi\alpha^\prime}\dot c_{-n}^iB^{-1}_{ij}
c_{-n}^j+nc_{-n}^ic_n^i\right)\right],
\end{eqnarray}
where $f_n\ne 0, ~ n>0$ are real numbers. While they can be removed by 
shift of the variables $c_n$, it is convenient to keep them in the 
action. Starting from any particular $S_f$, the variables $c_n$ can be 
taken as the ones which generate the string coordinate (\ref{4}), the 
latter will not depend on $f_n$. At the same time, brackets of $c_n$ 
and the Hamiltonian $H_f$ will depend on $f_n$, so there is appear 
natural arbitrariness in the induced bracket for $x^i(\tau, \sigma)$. 
Choice of some particular form of the bracket implies that the 
corresponding Hamiltonian $H_f$ must be associated with the Lagrangian 
formulation (\ref{1})-(\ref{3}). 

\noindent
{\bf Hamiltonian analysis of the machanical action.} 
Direct application of the Dirac algorithm to the action  (\ref{8}), gives us the primary second class constraints 
\begin{eqnarray}\label{9}
G_{0i}&\equiv& p_{0i}-\frac12B_{ij}c_0^j=0, \cr 
G_{ni}&\equiv& p_{ni}-\pi\alpha^\prime f_nB_{ij}c_n^j=0, \cr
G_{-ni}&\equiv& p_{-ni}+\frac{f_n}{4\pi\alpha^\prime}B^{-1}_{ij}c_{-n}^j=0.
\end{eqnarray}

\noindent Their Poisson brackets are

\begin{eqnarray}\label{10}
\{G_{0i}, G_{0j}\}&=&-B_{ij}, \cr
\{G_{ni}, G_{mj}\}&=&-2\pi\alpha^\prime f_nB_{ij}\delta_{n-m,0}, \cr
\{G_{-ni}, G_{-mj}\}&=&\frac{f_n}{2\pi\alpha^\prime}B^{-1}_{ij}
\delta_{n-m,0},
\end{eqnarray}
and the corresponding Hamiltonian is
\begin{eqnarray}\label{11}
H=&&\sum_{n=1}^\infty -f_nnc_{-n}c_n +
\lambda_0(p_0-\frac 12Bc_0)+ \nonumber \\
&&\lambda_n(p_n-\pi\alpha^\prime f_nBc_n)+ 
\lambda_{-n}(p_{-n}+\frac{f_n}{4\pi\alpha^\prime}B^{-1}c_{-n}).
\end{eqnarray}
There are no secondary constraints in the problem. From the 
consistency conditions that constraints do not evolve in time, $\dot G=0$,
one obtains expressions for the Lagrangian multipliers 
\begin{eqnarray}\label{12}
\lambda_0=0, \qquad \lambda_n=-\frac{n}{2\pi\alpha^\prime}c_{-n}B^{-1}, 
\qquad \lambda_{-n}=2\pi\alpha^\prime nc_nB.
\end{eqnarray}
To take into account the second class constraints (\ref{9}) one 
introduces the Dirac bracket 
\begin{eqnarray}\label{13}
\{K, P\}_D=&&\{K, P\}
+\{K, G_{0i}\}(B^{-1})^{ij}\{G_{0j}, P\}+ \nonumber \cr
&&\sum_{n=1}^\infty
\{K, G_{ni}\}\frac{1}{2\pi\alpha^\prime f_n}
(B^{-1})^{ij}\{G_{nj}, P\}-\nonumber \\
&&\{K, G_{-ni}\}\frac{2\pi\alpha^\prime}{f_n}B^{ij}
\{G_{-nj}, P\}.
\end{eqnarray}
After that, the variables $p_n^i$ can be omitted from consideration. 
The resulting Hamiltonian formulation for (\ref{8}) consist of the 
physical variables $c_n^i$ with the Dirac brackets 
\begin{eqnarray}\label{14}
\{c_0^i, c_0^j\}_D&=&-(B^{-1})^{ij}, \cr
\{c_n^i, c_m^j\}_D&=&-\frac{1}{2\pi\alpha^\prime f_n}
(B^{-1})^{ij}\delta_{n-m,0}, \cr
\{c_{-n}^i, c_{-m}^j\}_D&=&\frac{2\pi\alpha^\prime}{f_n}
B^{ij}\delta_{n-m,0}, \qquad n, m>0,
\end{eqnarray}
whose dynamics is governed now by the Hamiltonian 
\begin{eqnarray}\label{15}
H=-\sum_{n=1}^\infty f_nnc^i_{-n}c^i_n.
\end{eqnarray}
As it should be for the first order system, the Hamiltonian equations 
of motion which follow from (\ref{14}), (\ref{15}) are the equations 
(\ref{7}). Note also that the variables $c_n^i$ with $n$ fixed are 
canonically conjugated to each other. 
\newpage
The equations (\ref{7}) for the physical variables can be solved now 
in terms of oscillators 
\begin{eqnarray}\label{16}
c^i_n(\tau)&=&\alpha^i_ne^{in\tau}+\alpha^i_{-n}e^{-in\tau}, \cr
c^i_{-n}(\tau)&=&-2i\pi\alpha^\prime\alpha^i_nBe^{in\tau}+
2i\pi\alpha^\prime\alpha^i_{-n}Be^{-in\tau}, \cr
{\alpha^i_n}^* &=&\alpha^i_{-n}.
\end{eqnarray}
From Eq. (\ref{14}) one finds their brackets 
\begin{eqnarray}\label{17}
\{c_0^i, c_0^j\}_D&=&-(B^{-1})^{ij}, \cr
\{\alpha_n^i, \alpha_m^j\}_D&=&-\frac{1}{4\pi\alpha^\prime f_{\mid n\mid}}
(B^{-1})^{ij}\delta_{n+m,0}, \qquad n, m\ne 0, 
\end{eqnarray}
while the Hamiltonian (\ref{15}) acquires the form 
\begin{eqnarray}\label{18}
H_f=4i\pi\alpha^\prime\sum_{n=1}^\infty f_nn\alpha_n^iB_{ij}\alpha_{-n}^j.
\end{eqnarray}
It is worth nothing that this
procedure, being applied to the open string with the Neumann boundary
conditions, leads to the standard results [18, 19].

\noindent
{\bf Hamiltonian formulation for the open string with a B-field.}
By using of Eqs. (\ref{4}), (\ref{16}) we restore the string coordinate 
$x^i(\tau, \sigma)$ in terms of oscillators 
\begin{eqnarray}\label{19}
x^i(\tau, \sigma)=c_0^i+\sum_{n\ne 0}\frac{e^{in\tau}}{\mid n\mid}
\left(\alpha_n^i\cos n\sigma-
2i\pi\alpha^\prime\alpha_n^jB_j{}^i\sin n\sigma\right)  
\end{eqnarray}
Thus, with the system (\ref{1})-(\ref{3}) one associates the 
Hamiltonian formulation which includes the variables 
$c_0^i, ~ \alpha_n^i, ~ 
n\ne 0$ with the brackets (\ref{17}) and the Hamiltonian (\ref{18}).
As a consequence of Eqs. (\ref{17}), (\ref{19})  
the induced bracket of the string coordinates turns out to be 
noncommutative in the bulk and on the boundary 
\begin{eqnarray}\label{20}
\lbrace x^i(\tau, \sigma), x^j(\tau, \sigma^\prime\rbrace_D&=&-(B^{-1})^{ij}-  
\sum_{n=1}^\infty\frac{1}{n^2f_n}\left(\frac{1}{2\pi\alpha^\prime}(B^{1})^{ij} \right. \nonumber \\
&& \left. \times \cos n\sigma\cos n\sigma^\prime - 2\pi\alpha^\prime B^{ij}\right. \nonumber \\
&& \left. \times \sin n\sigma\sin n\stackrel{~}{\sigma^\prime}\right).
\end{eqnarray}
In particular, the coordinates on D-brane obey
\begin{eqnarray}\label{21}
\{x^i(\tau, 0), x^j(\tau, 0\}_D=&&
\{x^i(\tau, \pi), x^j(\tau, \pi\}_D \cr 
=&&-\left(1+\frac{1}{2\pi\alpha^\prime}
\sum_{n=1}^\infty\frac{1}{n^2f_n}\right)(B^{-1})^{ij}.
\end{eqnarray}
It is interesting to note that the standard normalization 
$f_n\sim\frac{1}{n}$ of the oscillator brackets (\ref{17}) implies 
that one needs to use some regularization for Eq. (\ref{21}). 
For the choice $f_n=-\frac{\pi}{12\alpha^\prime}$ the D-brane plane 
becomes commutative, and the Hamiltonian in this case is 
\begin{eqnarray}\label{22}
H=-\frac{i\pi^2}{3}\sum_{n=1}^\infty\alpha_n^iB_{ij}\alpha_{-n}^j.
\end{eqnarray}

\noindent
{\bf Final Discussions.}
Finally, let us discuss freedom in definition of phase space 
bracket in the Hamiltonian formalism. Transition from configuration 
to phase space description is not unique procedure since one needs 
to define simultaneously the bracket and the Hamiltonian. 
One possibility is to start from the Poisson bracket, then exact 
expression for the Hamiltonian is known for a general case [16, 20].
According to the Darboux theorem, one can equally take as the starting 
point any nondegenerated closed two-form and then try to define a 
Hamiltonian from the condition that corresponding equations of motion 
are equivalent to the initial ones. Turning to our case (\ref{14}), 
(\ref{15}) it will be sufficiently to consider the brackets 
\begin{eqnarray}\label{23}
\{c_0^i, c_0^j\}_D&=&-E^{ij}, \nonumber \\  
\{c_n^i, c_m^j\}_D&=&-\frac{1}{2\pi\alpha^\prime f_n}A^{ij}\delta_{n-m,0}, 
~ n, m>0,
\end{eqnarray}
with some antisymmetric nondegenerated constant matrices $E$ and $A$. 
From the condition that the dynamics (\ref{7}) is reproduced in this 
formulation one finds
\begin{eqnarray}\label{24}
\{c_{-n}^i, c_{-m}^j\}_D=\frac{2\pi\alpha^\prime}{f_n}(AB^2)^{ij}
\delta_{n-m,0}, \cr 
H=-\sum_{n=1}^\infty f_nnc_{-n}(AB)^{-1}c_n.
\end{eqnarray}
One can take Eqs. (\ref{23}), (\ref{24}) instead of Eqs. (\ref{14}), 
(\ref{15}) as the Hamiltonian formulation corresponding to the system 
(\ref{8}). Repeating the previous analysis, one finds the same 
expression (19) for the string coordinate in terms of oscillators 
which obey 
\begin{eqnarray}\label{25}
\{c_0^i, c_0^j\}_D&&=-E^{ij}, \nonumber \cr
\{\alpha_n^i, \alpha_m^j\}_D &&=-\frac{1}{8\pi\alpha^\prime f_{\mid n\mid}}
\left(A-sgn(nm)B^{-1}AB\right)^{ij} \nonumber \\
&& \times \delta_{\mid n\mid-\mid m\mid ,0},~ n, m\ne 0,
\end{eqnarray}
The quantities $f_n, ~ E, ~ A$ can be chosen in such a way 
that the noncommutativity parameter on D-brane acquires exactly the same 
form as it was obtained from the disk propagator [2]. Namely, for the 
choice 
\begin{eqnarray}\label{26}
f_n&=&\frac{\pi^2}{3(2\pi\alpha^\prime)^3}, \nonumber \\ 
E&=&\frac{(2\pi\alpha^\prime)^2}{2}A, \nonumber \\ 
A&=&\left(1+(2\pi\alpha^\prime)^2B^2\right)^{-1}B,
\end{eqnarray}
one obtains the following Hamiltonian formulation for the system 
(\ref{1})-(\ref{3})
\begin{eqnarray}\label{27}
H&=&\frac{i\pi^2}{3(2\pi\alpha^\prime)^2}\sum_{n=1}^\infty
n\alpha_nB^{-1}\left(1+(2\pi\alpha^\prime)^2B^2\right)\alpha_{-n}, \nonumber \cr
\{c_0^i, c_0^j\}_D&=&-\frac{(2\pi\alpha^\prime)^2}{2}  
\left[\left(1+(2\pi\alpha^\prime)^2B^2\right)^{-1}B\right]^{ij}, \nonumber \cr
\{\alpha_n^i, \alpha_m^j\}_D&=&-\frac{3(2\pi\alpha^\prime)^2}{2\pi^2}
\left[\left(1+(2\pi\alpha^\prime)^2B^2\right)^{-1}B\right]^{ij}\nonumber \\
&&\times \delta_{n+m,0}, ~ n, m\ne 0,
\end{eqnarray}
which gives for the coordinates on D-brane the desired expression 
\begin{eqnarray}\label{28}
\{x^i, x^j\}_D=-(2\pi\alpha^\prime)^2(1-2i\pi\alpha^\prime B)^{-1}
B(1+2i\pi\alpha^\prime B)^{-1}.
\end{eqnarray}
%
%
%


\begin{thebibliography}{nn}
\bibitem{} A. Connes, M. R. Douglas and A. Schwarz, 
JHEP {\bf 02} (1998) 003 [hep-th/9711162]
\bibitem{} N. Seiberg and E. Witten,
JHEP {\bf 09} (1999) 032 [hep-th/9908142].
\bibitem{} C. G. Callan, C. Lovelace, C. R. Nappi and S. A. Yost,
Nucl. Phys. {\bf B 288} (1987) 525; A. A. Abouelsaood,  
C. G. Callan, C. R. Nappi and S. A. Yost, 
Nucl. Phys. {\bf B 280} (1987) 599.
\bibitem{} H. Arfaei and M. M. Sheikh-Jabbari,
Nucl. Phys. {\bf B 526} (1998) 278. 
\bibitem{} M. M. Sheikh-Jabbari,  Phys. Lett. {\bf B 425} (1998) 48.
\bibitem{} E. S. Fradkin and A. A. Tseytlin, Phys. Lett. {\bf B 163} 
(1985) 123.
\bibitem{} V. Schomerus, JHEP {\bf 06} (1999) 030 [hep-th/9903205].
\bibitem{} F. Ardalan, H. Arfaei and M. M. Sheikh-Jabbari, 
{\em Mixed Branes and M(atrix) Theory on Noncommutative Torus}, 
hep-th/9803067; JHEP {\bf 02} (1999) 016 [hep-th/9810072]; 
Nucl. Phys. {\bf B 576} (2000) 578 [hep-th/9906161].
\bibitem{} C. S. Chu and P. M. Ho, 
Nucl. Phys. {\bf B 550} (1999) 151 [hep-th/9812219]; 
Nucl. Phys. {\bf B 568} (2000) 447 [hep-th/9906192].
\bibitem{} S. Girvin and T. Jach, Phys. Rev. {\bf D 29} (1984) 5617. 
\bibitem{} G. Dunne and R. Jackiw, Nucl. Phys. Proc. Suppl. 
{\bf 33 C} (1993) 114 [hep-th/9204057]. 
\bibitem{} D. Bigatti and L. Susskind, Phys. Rev. {\bf D 62} (2000)  
066004 [hep-th/9908056].
\bibitem{} T. Lee, Phys. Rev. {\bf D 62} (2000) 024022 [hep-th/9911140].
\bibitem{} W. T. Kim and J. J. Oh, Mod. Phys. Lett. {\bf A 15} (2000) 
1597 [hep-th/9911085].
\bibitem{} P. A. M. Dirac, {\em Lectures on Quantum Mechanics}, 
Yeshiva University, NY, 1964.
\bibitem{} D. M. Gitman and I. V. Tyutin, {\em Quantization of Fields
with Constraints}, Springer-Verlag, 1990.
\bibitem{} L. Brink and M. Henneaux, {\it{Principles of String Theory}},
Plenum Press, NY and London, 1988.
\bibitem{} M.B. Green, J.H. Schwarz and E. Witten,
{\it{Superstring Theory}}, v.1,2, Cambridge Univ. Press, Cambridge, 1987.
\bibitem{} J. Polchinski, {\it{String Theory}}, v.1,2,
Cambridge University Press, 1998.
\bibitem{} A. A. Deriglazov and K. E. Evdokimov, Int. J. Mod. Phys. 
{\bf A 15} (2000) 4045 [hep-th/9912179].
\end{thebibliography}
\end{document}